\documentclass[review]{elsarticle}

\usepackage{hyperref, graphicx}
\usepackage{float}
\journal{Preventive Medicine Report}

 \usepackage{numcompress}
 
\begin{document}

\begin{frontmatter}

\title{Interactive Network Visualization of Opioid Crisis Related Data- Policy, Pharmaceutical, Training, and More}


\author[label1]{Olga Scrivner\corref{mycorrespondingauthor}}\cortext[mycorrespondingauthor]{Corresponding author: obscrivn@indiana.edu}
\author[label2]{Elizabeth McAvoy}
\author[label2,label3]{Thuy Nguyen}
\author[label1]{Tenzin Choeden}
\author[label2]{Kosali Simon}
\author[label1]{Katy B{\"o}rner}

\address[label1]{Luddy School of Informatics, Computing, and Engineering, Indiana University}
\address[label2]{O'Neill School of Public and Environmental Affairs, Indiana University}
\address[label3]{School of Public Health, University of Michigan}

\begin{abstract}
Responding to the U.S. opioid crisis requires a holistic approach supported by evidence from linking and analyzing multiple data sources. This paper discusses how 20 available resources can be combined to answer pressing public health questions related to the crisis. It presents a network view based on U.S. geographical units and other standard concepts, crosswalked to communicate the coverage and interlinkage of these resources. These opioid-related datasets can be grouped by four themes: (1) drug prescriptions, (2) opioid related harms, (3) opioid treatment workforce, jobs, and training, and (4) drug policy. An interactive network visualization was created and is freely available online; it lets users explore key metadata, relevant scholarly works, and data interlinkages in support of informed decision making through data analysis. 

\end{abstract}

\begin{keyword}
Opioid Crisis \sep Policy \sep Network Visualization \sep Data Linkage


\end{keyword}

\end{frontmatter}


\section{Introduction}
\label{S1}

The  U.S. opioid epidemic is a major national concern, with the number of fatal drug overdoses accelerating during the COVID-19 pandemic. As of May 2020, the 12-month counts of reported deaths from drug overdose have increased by an estimated 17\% compared with the year 2019—rising from 67,281 to 79,251 deaths~\cite{Ahmad2020}. Furthermore, according to a recent study of spatial and temporal overdose spikes by the Overdose Detection Mapping Application Program, the number of reported overdoses has increased by 18\% between pre- (Jan 1 through  March 18, 2020) and post-stay-at-home order (March 19 through May 19, 2020), while the number of counties reporting fatalities has increased~\cite{Alter2020}. 

To address the current opioid crisis, the Department of Health and Human Services (HHS) strategic priorities include improvements in (1) pain management, (2) prevention, treatment and recovery, (3) data and research related to the opioid crisis, and (4) overdose-reversing drugs~\cite{Price2017}. A holistic understanding of multiple datasets of drug policy, pharmacy claims, treatment workforce, and opioid-related harms can advance research related to the opioid crisis. We focus our discussion on data resources that are available without major hurdles to access\footnote{We acknowledge that there are many valuable data resources. however they are harder to incorporate into research, due to privacy protection or cost concerns.}. These data often include aggregate-level identifiers, such as geographical units (state, county), drug names, and occupation codes. Using these aggregate-level identifiers can serve as linkages between datasets, and these linkages may allow researchers and stakeholders to identify new areas for public or health interventions and provide evidence-based guidelines for practitioners and patients. A systematic view of datasets suggests that data linkages become an ``informational asset'' transforming the way we observe and analyze data~\citep{weber2014}. Stakeholders and decision makers, however, are often challenged by the large number, complexity, and peculiarities of the existing datasets. Researchers may also not be aware of available resources as these datasets are provided by different sources and have varying data quality and coverage. Some datasets are freely available while others require signing of legal documents or payment of fees for additional aspects of the data. Furthermore, some datasets are massive in size requiring database expertise to run queries; other datasets exist only as textual data in a PDF format and require file parsing before usage. This review seeks to identify commonly used, freely available datasets and describe their coverage and linkages at the aggregate levels. This paper contributes to the growing body of work on linking data  sources~\cite{weber2014,Smart2018,smart2020,saloner2020} by introducing a novel visualization of linked data that communicates their temporal, geospatial, topical coverage, and highlights the interlinkages between them. Researchers and practitioners can use this visualization to identify datasets relevant for research, teaching, and policy decisions. 

\section{Background}
The causes, consequences, and manifestations of the U.S. opioid crisis have been studied from many different angles, including prevention, treatment, drug prescription, law enforcement, criminal justice, and overdose reversal.  
Treatment expansions and 
prescription reductions are two essential steps in reducing mortality and improving safety for patients with chronic pain. Monitoring and regulatory policies play an equally important role in balancing between harms, cost, availability, and benefits of opioid use, as seen in policies such as prescription drug monitoring programs (PDMPs), health insurance expansions, and comprehensive federal legislation (e.g., the Comprehensive Addiction and Care Act)~\cite{Poitras2018,Scrivner2020}. These efforts have led to a decrease in the overall U.S. drug prescription rate, which has fallen from 81.3 per 100 people in 2012 to 46.7 in 2019~\cite{CDC2019}
. But while the U.S. has had success in implementing preventative measures, it has struggled with improving treatment access for those suffering from addiction. A major gap remains between service demand and supply: only 30\% of U.S. adults with Opioid Use Disorder (OUD) have reported receiving a treatment, according to the 2015-2017 National Survey on Drug Use and Health data (NSDUH). In addition, the 2017 Treatment Episode Dataset (TEDs) reveals an increase in Opiate-related admissions (682,074), whereas only 1,691 out of 15,961 treatment facilities are OTP certified (see the 2019 National Survey of Substance Abuse Treatment Services [N-SSATs]). In terms of the number of establishments, the 2018 County Business Patterns data (CBP) identifies 46,254 Substance Use Disorder Treatment (SUDT) outpatient centers, 42,906 Residential SUDT facilities, and 692 SUDT hospitals.  Despite the high priority for training expressed by the U.S. Department for Health and Human Service and high job demand, the behavioral health workforce (integrated mental and substance use disorder) has been ``characterized as being in crisis''~\cite[p. 15]{Skillman2016}. The interdependence of these social, health, economic, and public policy factors calls for an interdisciplinary holistic and systematic approach where researchers and practitioners can zoom out and examine the problem as a whole and then zoom in to solve the most pressing issues that have the highest positive impact on improving health and services while decreasing crime and addictions-related disorders.

Recently, several studies were published that review the current literature and secondary data relevant to the opioid addiction crisis~\cite{leece2019,smart2020,Maclean2020}. Maclean et al.~\citep{Maclean2020} collected and reviewed economic studies and identified several topics relevant for understanding the opioid crisis: (1) pharmaceutical industries and drug prescriptions, (2) healthcare providers and labor market, (3) harms and crime, (4) policies. Another study, ~\cite{leece2019} 
extracted intervention variables (e.g., prevention, treatment, harm reductions) and enabling variables (e.g., surveillance, stigma). Furthermore, Smart et al.~\cite{Smart2018,smart2020} reviewed existing datasets, grouping them according to the HHS strategic priorities: (1) better pain management, (2) addiction prevention, treatment and recovery service, and (3) better targeting of overdose-reversing drugs. In addition, authors classified data based on type, 
namely national surveys, electronic health records (EHR), claims data, mortality records, prescription monitoring data, contextual and policy data, and others (national, state, local). 
Strengths and weaknesses of each dataset were assessed using various metrics (e.g., data accessibility, data linkage, coverage).

Data descriptions are often presented in a tabular format with new attributes rendered as columns.  For instance, in~\cite{leece2019}, each variable is provided with its relative frequency of occurrence in the reviewed literature, whereas in~\cite{smart2020}, a plus/minus sign is used to indicate strengths and weaknesses for each dataset. A different perspective, called ``probabilistic linkage,'' was developed by Weber et al.~\cite{weber2014} in 2014 focusing on a visual representation of potential biomedical sources and the values of their linkages. The team used a tabular form with sizes, shapes, colors, and positions to indicate data quality, data linkage, types of data (e.g., pharma, claims, EHR, non-clinical data), data coverage, and even the probabilities for obtaining new data or linking existing data.

Over the last several years, many new datasets became available (e.g., data.gov and nlm.nih.gov), and researchers now have access to datasets with diverse quality and coverage. In order to federate and use these resources, detailed knowledge about the datasets is required. Understanding data linkages~\cite{Shlomo2018} becomes critical for understanding, communicating, and reducing disease~\cite{neish2015}. Data visualizations can be used to communicate the complexity of heterogeneous data. For example, SPOKE~\cite{Baranzini2021} and Springer Nature SciGraph~\cite{SpringerNature2020} use a knowledge graph (KG) to interlink and query different datasets. The SPOKE KG interlinks more than 30 publicly available biomedical databases, whereas SciGraph interlinks funders, projects, publications, citations, and scholarly metadata in support of data exploration.

\section{Methods}
\subsection{Data Collection}

\begin{figure}[H]
    \centering
    \includegraphics[height=3.1in]{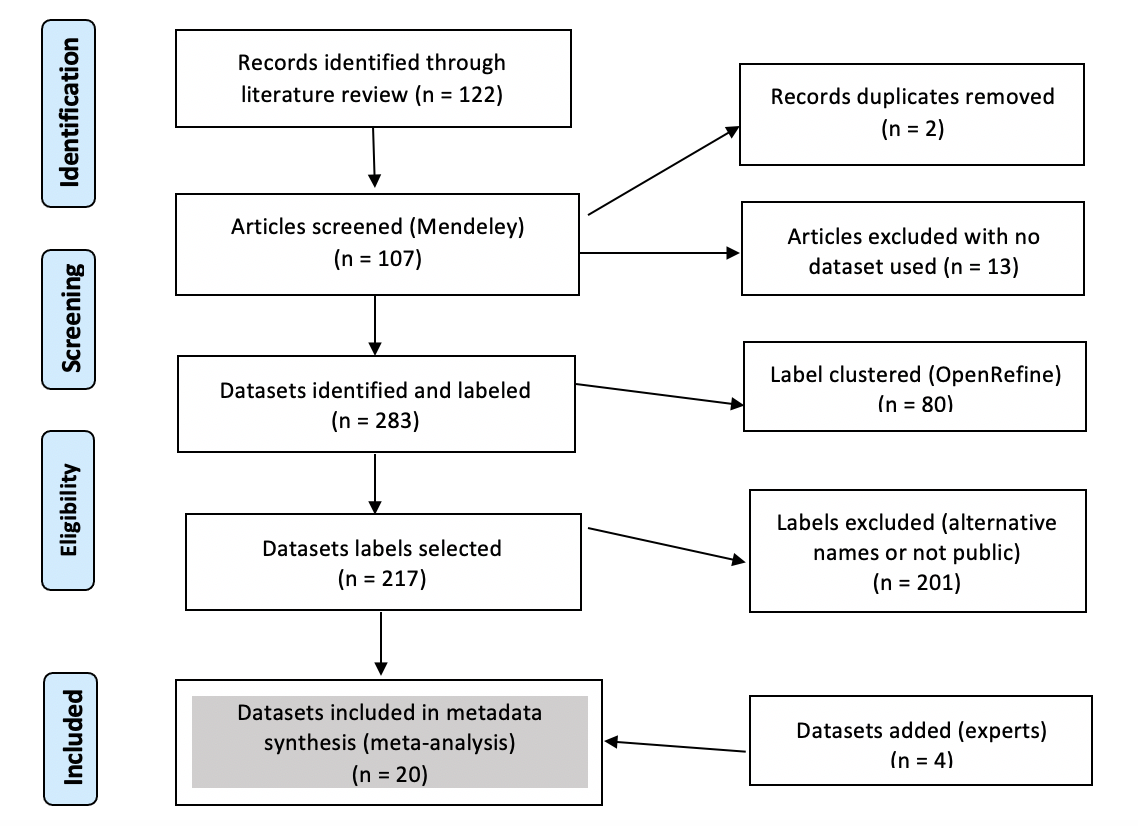}
    \caption{PRISMA flow diagram of the scoping review process}
    \label{fig:fig1}
\end{figure}

A recent review of the economics literature related to the opioid crisis by Maclean et al. described over 100 
major economic studies on the U.S. opioid crisis based on a comprehensive review of the literature and expert consultations~\cite{Maclean2020}. 
Building on this work, we applied a modified protocol of scoping reviews~\cite{arksey2005} to identify open datasets used in the 120 studies cited (see Figure~\ref{fig:fig1}). 
Specifically, the 120 articles ranging from 1986 to 2020 were imported to the Mendeley library group, and duplicate records were removed.
Each article was scanned for datasets mentioned in the methodology section and articles without datasets were discarded. The remaining set (107 articles) was tagged in Mendeley with dataset names as they were used in the studies. We identified 283 unique name tags.
Across the 107 studies, there were many inconsistencies in naming and spelling, for instance, `nvss,' `nvss multiple cause of death,' and `nvss multiple cause-of-death mortality' all referred to U.S mortality data from death certificates, produced by the National Center for Health Statistics. We normalized labels using OpenRefine and the Nearest Neighbor algorithm with Prediction by Partial Matching (PPM) distance~\cite{Stephens2018}. The algorithm detected 61 clusters that were merged, resulting in 230 normalized labels. 
We manually inspected all labels and removed datasets that did not fit our eligibility criteria: (1) dataset must be publicly available, and (2) dataset should fall into one of the following categories: i) pharmaceutical data--related to opioid prescription, ii) policy data--related to state drug laws, iii) opioid overdose data--related to treatment and treatment results, and iv) employment data--related to training and hiring in the substance use disorder treatment industry (SUDT). As a result, we identified 20 datasets for synthesis and data linkage exploration (see Table 1). 

\begin{table}[H]
    \centering
    \caption{Datasets to support research on the opioid crisis. 16 datasets are extracted from the review study~\cite{Maclean2020} while 4 additional datasets underlined are ones we identified as relevant. 
    Datasets marked with * require a request submission prior to download.}
    \begin{tabular}{l|ll}
     \textbf{Dataset} & \textbf{Description} &\textbf{Type} \\
     \hline
     \hline
       CDC Mortality &	CDC Opioid Overdose Rate & Harms\\
TEDS-A &Treatment Episode Data Set: Admissions & Harms\\
\underline{NAS}* &	National Alcohol Survey & Harms \\
NSDUH &	National Survey on Drug Use and Health & Harms \\
\underline{NPDS}* &	National Poison Data System & Harms \\
\underline{TEDS-D} & Treatment Episode Data Set: Discharge & Harms \\
N-SSATs &	National Survey of Substance Abuse Treatment Services & Jobs \\
QCEW &	Quarterly Census of Employment and Wages & Jobs\\
\underline{IPEDS} &	Integrated Postsecondary Education Data System & Jobs \\
CBP	& County Business Patterns & Jobs\\
ACS	& American Community Survey & Jobs\\
MEPS &	Medical Expenditure Panel Survey & Pharma\\
Sunshine Act &	Open Payments & Pharma\\
SDUD (Medicaid) &	State Drug Utilization Data & Pharma \\
Medicare* &	Medicare Part D Prescription Drug Event & Pharma\\ 
ARCOS* &	Automated Reports and Consolidated Ordering System & Pharma \\
CDC Prescription &	CDC Drug Prescription & Pharma \\
PDMP &	Prescription Drug Monitoring Program & Pharma \\
PDAPS &	Prescription Drug Abuse Policy System & Policy \\
NAMSDL &	National Alliance for Model State Drug Laws & Policy\\
\hline
\hline

    \end{tabular}
    
    \label{tab:tab1}
    
\end{table}

\subsection{Data Analysis}
Data synthesis follows a 3-step process for each dataset: (1) data description (dictionary, size, and time coverage), (2) data linkages, and (3) scholarly metadata (relevant publications). For each dataset, we searched for a data download link and dictionary, which provides valuable information about data content and format. For some datasets, the dictionary URLs were not available. As a result, we provide this information only for 10 of 20 datasets in this study. 
Size was determined as the number of records 
based on the most recent year: (1) Small - less than 10,000, (2) Medium-sized - between 10,000 and 1,000,000, (3) Large - 1,000,000 or greater. Time coverage provides information on the year when the dataset became available and the most recent data available for download. Several data attributes are used to identify data linkages: geographical units (e.g., state, county) and standard crosswalks (e.g., the North American Industry Classification System or NAICS, Drug Name). Finally, we identified three recent publications that use a dataset to illustrate research results derived from that data. In total, 16 variables exist for each dataset: common abbreviation, full name, data description, dataset category, source URL (some missing data), 
dictionary URL, the number of records per year (most recent), size, time coverage (year start and year end), size, geo units, crosswalks, and three publications. 

\subsection{Network Visualization}  
Network visualizations are widely used to capture the relationship between entities (e.g., co-authorship network or gene-disease networks). They display entities (nodes) and their relationship (edges) in layouts that showcase overall connectivity structure and clusters while avoiding edge crossings. Networks can be extracted from tabular data, e.g., a co-author network can be extracted from a tabulation of papers and the set of authors per paper—co-author links connect all authors that appear in a paper together, creating an undirected weighted 
network~\cite{Emmert-Streib2018}. In addition, each node and edge can be color- or size-coded to visualize additional attributes (e.g., number of papers, number of citations, year of first publication, publication, topic). 

\begin{figure}
    \centering
    \includegraphics[height=0.8in]{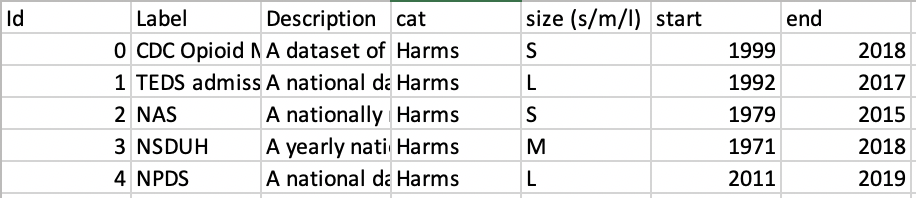} \includegraphics[height=0.78in]{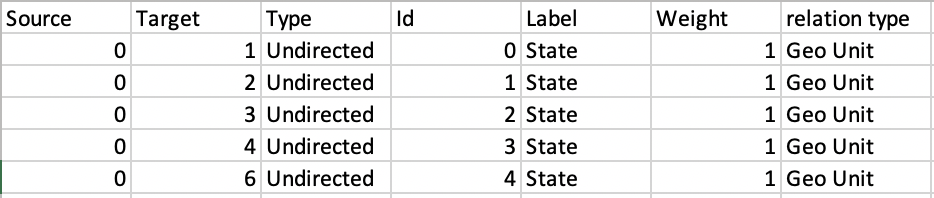}
    \caption{Nodelist, partial—only 7 of 16 attributes are shown (top) and Edgelist 
    (bottom)}
    \label{fig:fig2}
\end{figure}

To compute a visualization of the 20 datasets, we first converted the csv file  with all 20 datasets (rows) and 16 attributes (columns) into two separate files, namely nodelist and edgelist. The nodelist has an additional numeric identifier for each dataset that is used in the edgelist to describe how datasets are linked. For instance,  the ID for CDC Mortality dataset is `0' and the ID for TEDS admission is `1' (see Figure~\ref{fig:fig2}). These two datasets share the same attribute `State.' Thus, we can build their linkage from CDC Mortality (source) to TEDs admission (target) and vice versa, since the network is undirected. 
The resulting network has 20 nodes of four categories 
and 117 edges of xx different types.

We used the Force Atlas 2 algorithm in Gephi~\cite{Bastian2009} to layout the network in a 2-dimensional space in a manner that minimizes edge crossings and stress: i.e., interlinked nodes are in close proximity (see Figure 2). Datasets are color-coded to visually render 4 themes: 
prescription,  harms,  jobs, and policy. 
The workflow for creating this network in Gephi is available at GitHub (https://github.com/obscrivn/datasets).
\begin{figure}[H]
    \centering
    \includegraphics[height=3in]{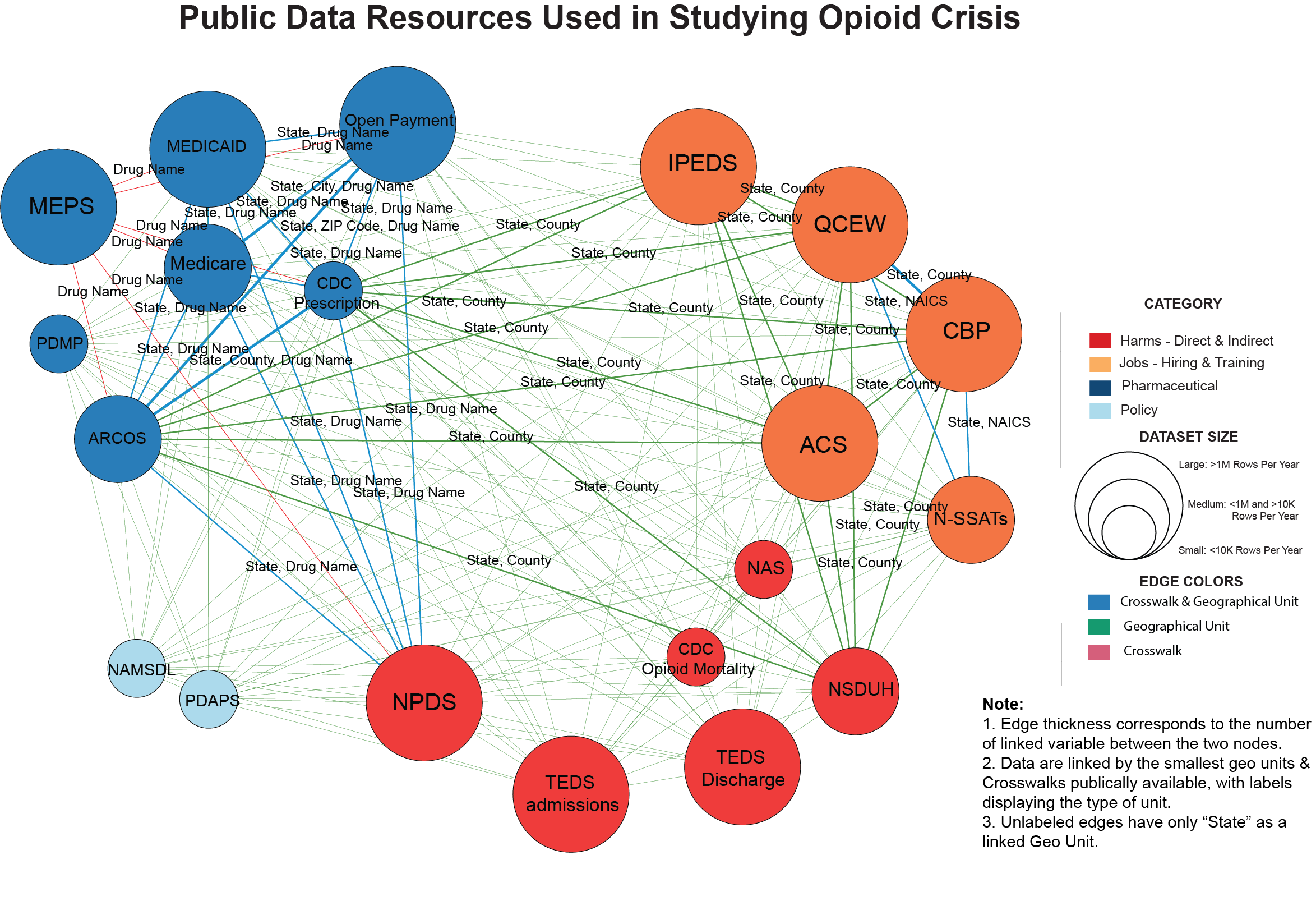}
    \caption{Network representation of the 20 datasets  with policy data (in light blue), pharmaceutical data (dark blue), opioid data (red), and jobs (hiring/training) data (orange). 
    Circle size corresponds to the size of the dataset. Edge color denotes the type of linkage.}
    \label{fig:fig3}
\end{figure}

\begin{figure}
    \centering
    \includegraphics[height=2.3in]{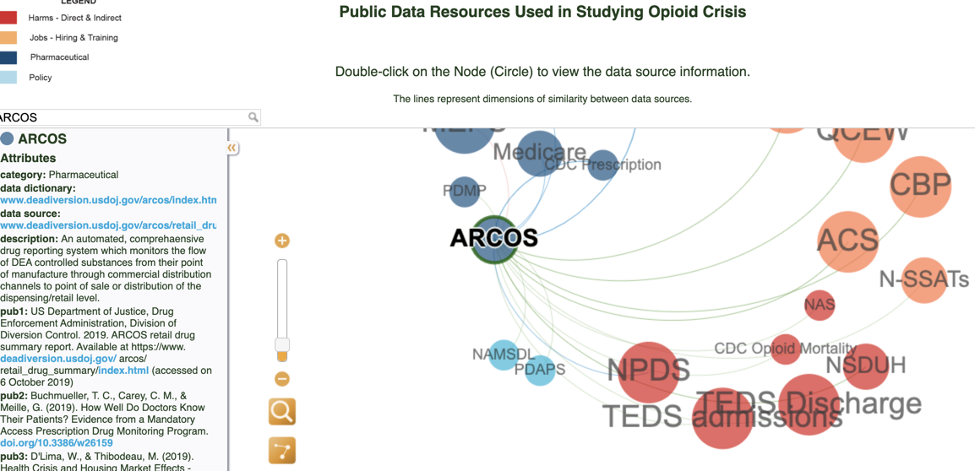}
    \caption{Interactive network visualization with legend in top left explaining color and size coding; details on demand in lower left; interactive network layout on right.}
    \label{fig:fig4}
\end{figure}

The interactive visualization is created using JavaScript GEXF viewer package~\cite{Velt2019}. The Gephi network is exported from Gephi into a gexf format (.gexf), a native xml format suitable for JavaScript (js) interactive visualization frameworks. Then, gefx.js code is updated and uploaded to GitHub. The interactive solution is available at https://obscrivn.github.io/datasets/ 
and it supports search, filter, and details on demand~\cite{Shneiderman1996}, as illustrated in Figure~\ref{fig:fig4}.
\section{Results and Discussion}
 
The visualization makes it possible to interactively explore key metadata and data interlinkages. Using the online site, users can explore and navigate each dataset by clicking on the node, examining the linked datasets, reviewing the data dictionary, and getting familiar with recent publications using the selected dataset. The collection of relevant scholarly articles helps researchers become familiar with case studies that use these datasets. Figure~\ref{fig:fig4} zooms into the ARCOS dataset.
The dark blue color in the legend specifies that this dataset belongs to a pharmaceutical category. 
By clicking on the ARCOS node, the attribute menu is shown on the left. The dictionary and dataset attributes provide direct links for reviewing the data dictionary and downloading data. From the data description, a potential user learns that this dataset provides information on drug sales and distribution. To provide relevant information about data usage, the visualization shows three recent scholarly publications using the ARCOS dataset. For instance, the study by~\cite{DLima2019} presents new evidence that changes in house prices near drug dispensaries are negatively correlated with drug quantities.  
A user might then check the ACS dataset, which is the American Community Survey about households. 
Next, the researcher can view time coverage and size for ARCOS: it ranges from 2009 to 2019 and the size of the dataset is medium. In addition, the menu specifies to which datasets ARCOS can be linked. For example, CDC Mortality and TEDS-admission share the 'State' attribute, whereas Open Payments, MEPS, and CDC Prescription share `Drug Name' attribute. The researcher can explore various hypotheses based on these potential links; for instance, \textit{Do states with high hospital admission rates and high prescription rates have also evidence of large payments to medical practitioners and some negative changes in households?}

\section{Conclusion}
One key priority laid out by HHS for combating the opioid crisis is better access to data and the encouragement of data-driven (policy) decision making. To assist researchers and policymakers navigating through existing datasets,
we developed a dataset and visualization that makes it possible to explore important characteristics and interlinkages of 20 widely used, publicly available datasets. Going forward, we plan to apply the same methodology to individual-level linked data and non-public resources. A current limitation of the presented work is the fact that the datasets are not updated as new data becomes available. In future research, we will perform regular updates of the datasets and their interlinkages. Another important area for future work is conducting  user studies to identify how to best improve the visualization for different stakeholder groups and what additional datasets should be added. 

\bibliography{mybibfile}

\end{document}